# New superconduting cuprates with no effective doping: T'-La$^{3+}_{2-x}$RE$^{3+}_x$CuO$_4$

Akio Tsukada[*], Yoshiharu Krockenberger, Hideki Yamamoto, Michio Naito

NTT Basic Research Laboratories, NTT Corporation, 3-1 Wakamiya, Morinosato, Atsugi-shi, Kanagawa 243-0198, Japan

**Abstract**

We report the synthesis of new superconducting cuprates T'-La$_{2-x}$RE$_x$CuO$_4$ (RE = Sm, Eu, Tb Lu, and Y) using molecular beam epitaxy. The new superconductors have no effective dopant, at least nominally. The substitution of isovalent RE for La was essentially performed to stabilize the T' phase of La$_2$CuO$_4$ instead of the T phase. The maximum $T_c^{onset}$ is ~ 25 K and $T_c^{zero}$ is ~ 21 K. The keys to our discovery are (1) the preparation of high-crystalline-quality La-based T' films by low-temperature (~ 650°C) thin film processes, and (2) more thorough removal of impurity oxygen at the apical site, which is achieved by the larger in-plane lattice constant ($a_0$) of T'-La$_{2-x}$RE$_x$CuO$_4$ than other T'-Ln$_2$CuO$_4$ (Ln = Pr, Nd, Sm, Eu, Gd) with the aid of large surface-to-volume ratio of thin films.

* Corresponding author

Dr. Akio Tsukada

Postal address: Superconducting Thin Films Research Group, NTT Basic Research Labs., 3-1, Wakamiya Morinosato, Atsugi-shi, Kanagawa 243-0198, Japan

Phone: +81-46-240-3349

Fax: +81-46-240-4717

E-mail address: tsukada@will.brl.ntt.co.jp

# 1. Introduction

Superconductivity in T' cuprates is very sensitive to impurity oxygen at the O(3) site (apical site). As-grown T' cuprates ($Ln_{2-x}Ce_xCuO_4$: $Ln$ = lanthanide element) are not superconducting [1, 2] because they contain a fair amount of impurity oxygen atoms at the apical site. Superconductivity appears only after heat treatment to remove apical oxygen. In T' superconductors, the superconducting (SC) region is adjacent to the antiferromagnetic (AF) region, and the superconductivity suddenly appears at the SC-AF boundary with maximum $T_c$ [3], suggesting competition between AF and SC orders. In the "optimally" doped $Pr_{1.85}Ce_{0.15}CuO_4$ or $Nd_{1.85}Ce_{0.15}CuO_4$, the AF correlations exist in as-grown non-superconducting samples, but they essentially disappear in reduced superconducting samples [4, 5]. This implies that more complete removal of apical oxygen would weaken the AF correlations and thereby expand the superconducting region. This was actually demonstrated by Brinkmann *et al.*, who achieved an extended superconducting region ($0.03 < x < 0.17$) in $Pr_{2-x}Ce_xCuO_4$ using an improved reduction technique [6].

In $La_{2-x}Ce_xCuO_4$, which can be stabilized by thin film synthesis [7], the superconducting region extend to lower doping ($0.45 < x < 0.22$) as compared with $Pr_{2-x}Ce_xCuO_4$ and $Nd_{2-x}Ce_xCuO_4$. Although superconductivity does not appear at $x = 0$, the end-member T'-$La_2CuO_4$ shows low resistivity (2 mΩcm at 300 K) and metallic behavior down to 150 K [8]. Our further efforts to improve metallicity toward a sign of superconducitivty in T'-$La_2CuO_4$ films were hampered by the upper limit of ~ 600°C in the synthesis temperature, which resulted in rather poor crystallinity. To see if we could stabilize the T' phase of $La_2CuO_4$ at > 600°C to improve crystallinity, we substituted large $La^{3+}$ by a small amount of small $RE^{3+}$ ($RE$ = rare earth element). This led to the discovery of superconducting T'-$La_{2-x}RE_xCuO_4$ with $T_c$ ~ 20 - 25 K. The reason for the superconductivity seems to be that the large in-plane lattice constant ($a_0$) of T'-$La_{2-x}RE_xCuO_4$ enables more thorough removal of impurity oxygen at the apical site with the aid of large surface-to-volume ratio of thin films.

## 2. Experimental

We grew $La_{2-x}RE_xCuO_4$ (*RE* = Pr, Nd, Sm, Eu, Tb, Lu, and Y) thin films in a customer-designed molecular beam epitaxy chamber from metal sources using multiple electron-gun evaporators [9]. The stoichiometry was adjusted by controlling the evaporation beam flux of each element using electron impact emission spectrometry via a feedback loop to the electron guns. During growth, 1 ~ 5 sccm of ozone gas (10% $O_3$ concentration) was supplied to the substrate for oxidation. The substrate temperature was typically ~ 650°C. The growth rate was ~ 1.5 Å/s, and the film thickness was typically ~ 900 Å. After the growth, the films were held at ~ 630°C in vacuum ($P_{O_2} < 10^{-8}$ Torr) for 10 minutes to remove interstitial apical oxygen. We mainly used $SrTiO_3$ (100) substrates but sometimes used $KTaO_3$ (100), $NdCaAlO_4$ (001), or $YAlO_3$ (100) substrates for the T' stabilization [8]. The lattice parameters and crystal structures of the grown films were determined using a standard X-ray diffractometer. Resistivity was measured by the standard four-probe method using electrodes formed by Ag evaporation.

## 3. Results and discussion

Figure 1(a) shows the temperature (*T*) dependence of resistivity ($\rho$) for the films of $La_{2-x}Tb_xCuO_4$ with different *x*. Metallic behavior ($d\rho/dT > 0$) is observed at *x*

= 0.15 - 0.75.  Superconductivity appears at $x$ = 0.15 - 0.45.  Above $x$ = 0.3, the resistivity increases with increasing $x$.  Figure 1(b) shows the $c$-axis lattice constant ($c_0$) of La$_{2-x}$Tb$_x$CuO$_4$ films as a function of Tb concentration $x$.  At $x$ < 0.09, the films stabilize in the T phase.  At $x$ = 0.09, the film is a two-phase mixture of T and T'.  A further substitution of Tb results in a single T' phase.  In the T' phase, $c_0$ monotonically decreases with increasing $x$, which indicates that the substitution of La by Tb in La$_{2-x}$Tb$_x$CuO$_4$ is successful up to $x$ = 2.  The $c_0$ agrees with the reported value ($c_0$ ~ 11.82 Å) for bulk Tb$_2$CuO$_4$ at $x$ = 2 [10].  This result indicates that Tb ionic size is close to Tb$^{3+}$ (1.04 Å) rather than Tb$^{4+}$ (0.88 Å) in the La$_{2-x}$Tb$_x$CuO$_4$ films, although Tb could take tetravalent state as well as trivalent state.  X-ray photoelectron spectroscopy (XPS) results also support the trivalent state.  Figure 2(a) shows the *in-situ* Tb-4$d$ XP spectrum of La$_{1.7}$Tb$_{0.3}$CuO$_4$ film, and Figs. 2(b) - (d) show the reference spectra of TbO$_2$, Tb$_4$O$_7$, and Tb$_2$O$_3$, included for comparison [11].  Spectrum (a) mostly coincides with spectrum (d), indicating that Tb valence is 3+ or close to 3+.

      Similar investigations have been performed on other trivalent *RE* ions (Pr, Nd, Sm, Eu, Lu, and Y), and superconductivity was observed except for Pr and Nd substitution [12].  As an example, Fig. 3 shows the $\rho$-$T$ curve for a La$_{1.85}$Y$_{0.15}$CuO$_4$ film with $T_c$ ~ 17 K.  The superconductivity is also confirmed by substantial diamagnetic signal in magnetization measurement with the magnetic field parallel to the film surface (Fig. 3 inset).  Figure 4 summarizes the *RE* doping dependence of $T_c$ in T'-La$_{2-x}$*RE*$_x$CuO$_4$.  Here, $T_c$ is defined as the temperature of zero resistance.  One can see two trends in the figure: (1) the $T_c$ increases with decreasing $x$ until the T-to-T' boundary is encountered, and (2) the superconducting region becomes wider as the ionic radius of *RE* increases from Lu to Sm.

Finally, we discuss the origin of superconducting carriers in T'-$La_{2-x}RE_xCuO_4$, where one can think of two possible scenarios. One is that, the oxygen deficiencies at the O(2) site are a source of effective electron carriers. Although quantitative occupancy is not known because there is no established technique for determining the site-specific occupancy of oxygen in thin films, neutron diffraction experiments performed on bulk T'-$Nd_2CuO_4$ or T'-$(Nd,Ce)_2CuO_4$ do not support the change of occupancy at the O(2) site between before and after heat treatment to remove apical oxygen [13-15]. The second scenario is that, the T'-$La_2CuO_4$ is not a Mott insulator, and has intrinsic carriers, as predicted by the band calculation for $Nd_2CuO_4$ [16]. This scenario is supported by our *in-situ* valence band photoemission results, where "non-doped" $La_{1.85}Eu_{0.15}CuO_4$ [12] ($T_c^{zero}$ = 20 K) film has been compared with "doped" $La_{1.9}Ce_{0.1}CuO_4$ ($T_c^{zero}$ = 28 K) film [17]. Both spectra showed a clear Fermi edge, and were essentially identical except for a rigid-band shift of ~ 0.2 eV. The spectra were in good agreement with the calculated density of states for $Nd_2CuO_4$ [16]. In order to clarify which scenario is the more probable, one has to investigate the size of Fermi surface with changing Ce concentration systematically. Such a study with angle-resolved photo emission spectroscopy is currently underway [17].

## 4. Summary

We prepared new superconducting T'-$La_{2-x}RE_xCuO_4$ ($RE$ = Sm, Eu, Tb, Lu, and Y) thin films with no effective dopant, at least nominally. The maximum $T_c^{onset}$ of ~ 25 K and $T_c^{zero}$ of ~ 21 K are obtained at the T-T' phase boundary. The larger $a_0$ of T'-$La_{2-x}RE_xCuO_4$ than that of other T'-$Ln_2CuO_4$ enables more thorough removal of impurity oxygen at the apical site, which leads to the superconductivity. Our results suggest that the end member compounds of T'-$Ln_2CuO_4$ are not Mott insulators.

**Acknowledgments**

The authors thank Dr. T. Yamada, Dr. A. Matsuda, Dr. H. Sato, Dr. S. Karimoto, Dr. K. Ueda, and Dr. J. Kurian for helpful discussions, and Dr. M. Morita, Dr. H. Takayanagi, and Dr. S. Ishihara for their support and encouragement throughout the course of this study.

**Figure captions**

Figure 1    (a) Temperature dependence of resistivity for $La_{2-x}Tb_xCuO_4$ films with different $x$.   Superconductivity appears at $x = 0.15 - 0.45$.   (b) $c$-axis lattice constant ($c_0$) of $La_{2-x}Tb_xCuO_4$ films as a function of Tb concentration.   Square and circles represent T and T'-phases, respectively.   The gray area at $x < 0.09$ indicates the region where the T phase is stable.   The symbols connected by vertical dotted line indicate multiple-phase formation.

Figure 2    *In-situ* Tb-4$d$ XPS spectrum of (a) $La_{1.7}Tb_{0.3}CuO_4$ thin film.   Spectra (b) $TbO_2$, (c) $Tb_4O_7$ and (d) $Tb_2O_3$ are for comparison [11].

Figure 3    $\rho$-T curve for $La_{1.85}Y_{0.15}CuO_4$ thin film.   Inset shows the result of magnetic measurement for the same sample.

Figure 4    The *RE* concentration dependence of $T_c$ for $La_{2-x}RE_xCuO_4$ films.   $T_c$ is defined as the temperature of zero resistance (closed circles: Sm, closed squares: Eu, closed triangles: Tb, open squares: Lu, open circles: Y).

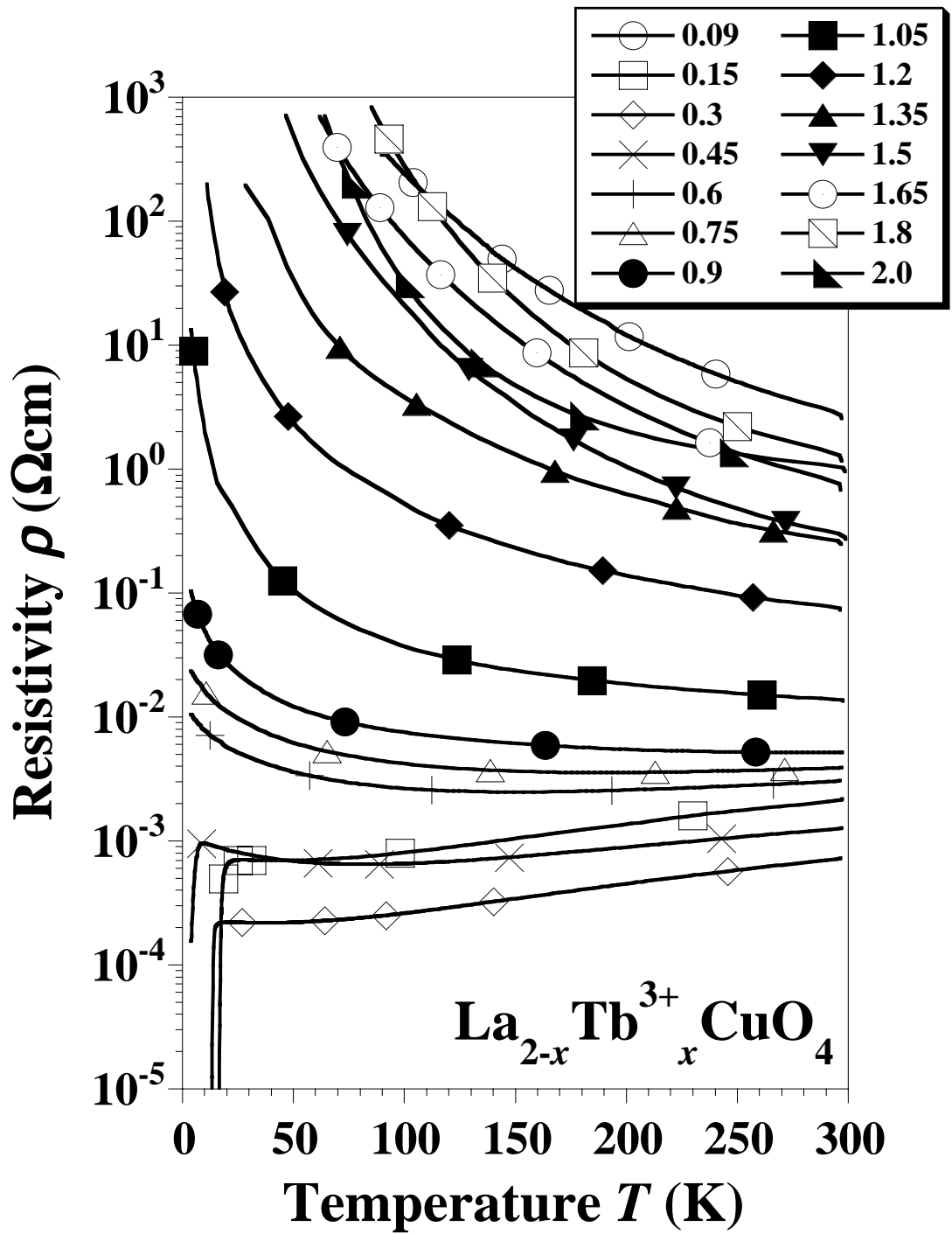

Figure. 1(a)  A. Tsukada *et al*.

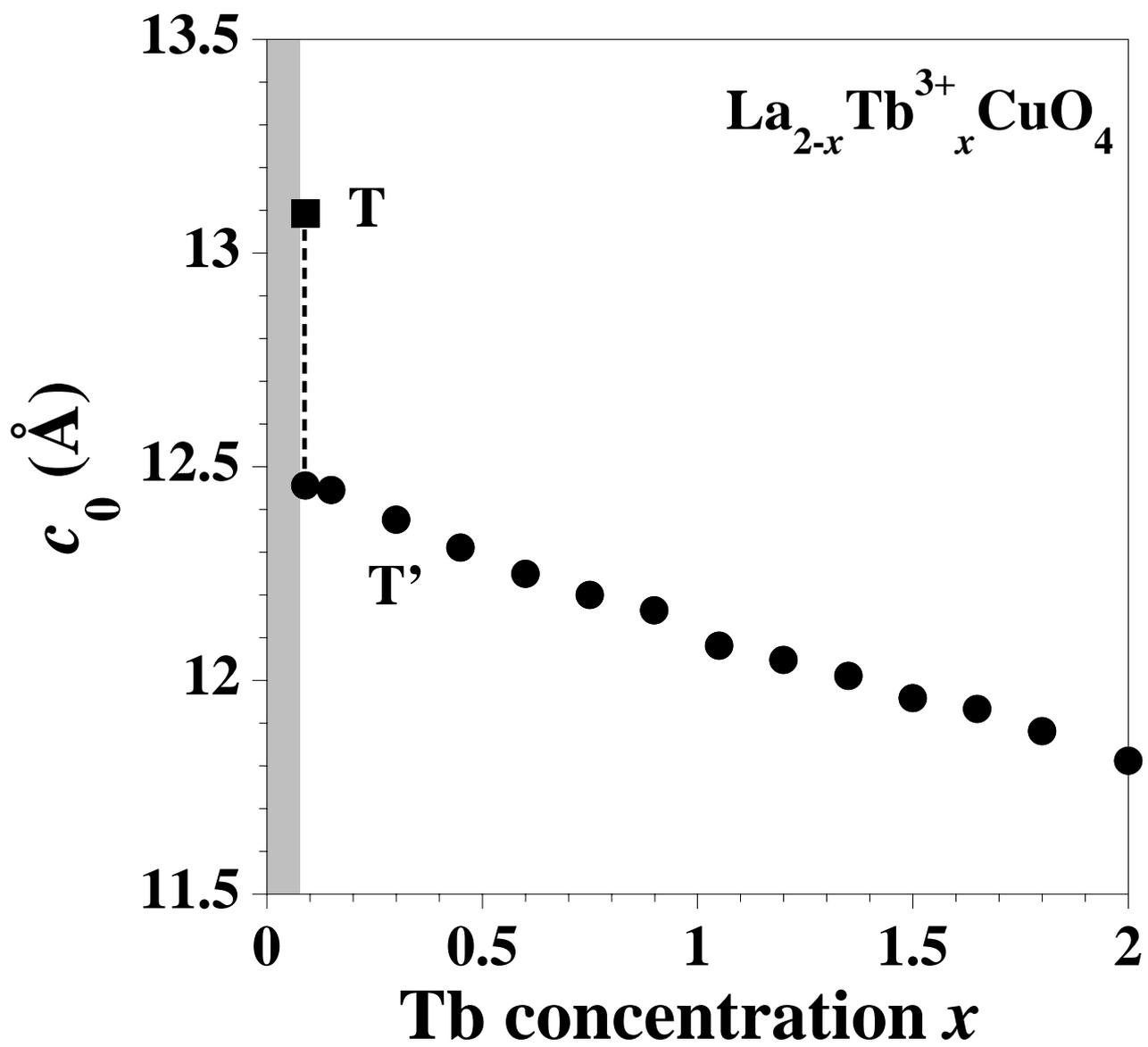

Figure. 1(b) A. Tsukada et al.

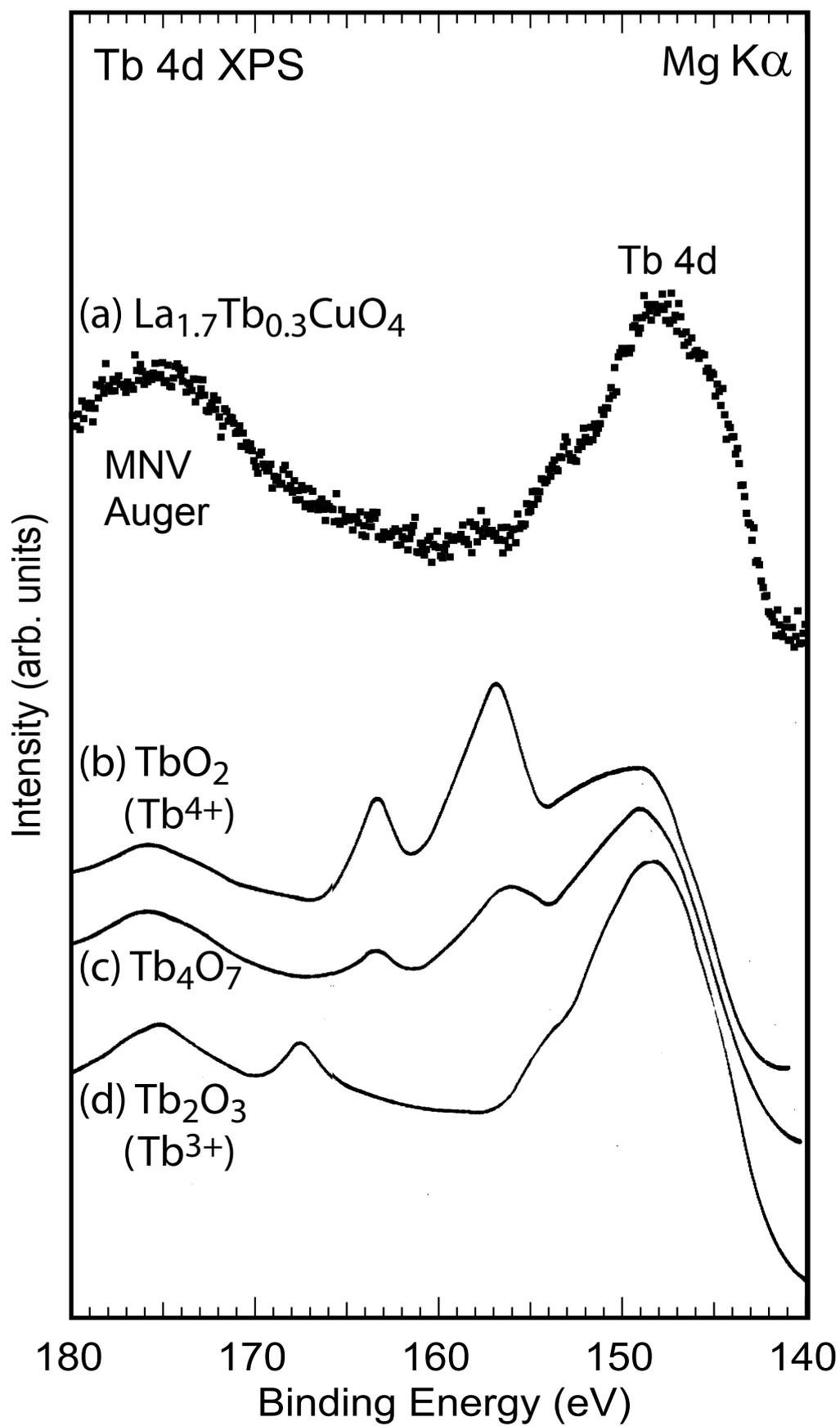

Figure 2. A. Tsukada *et al.*

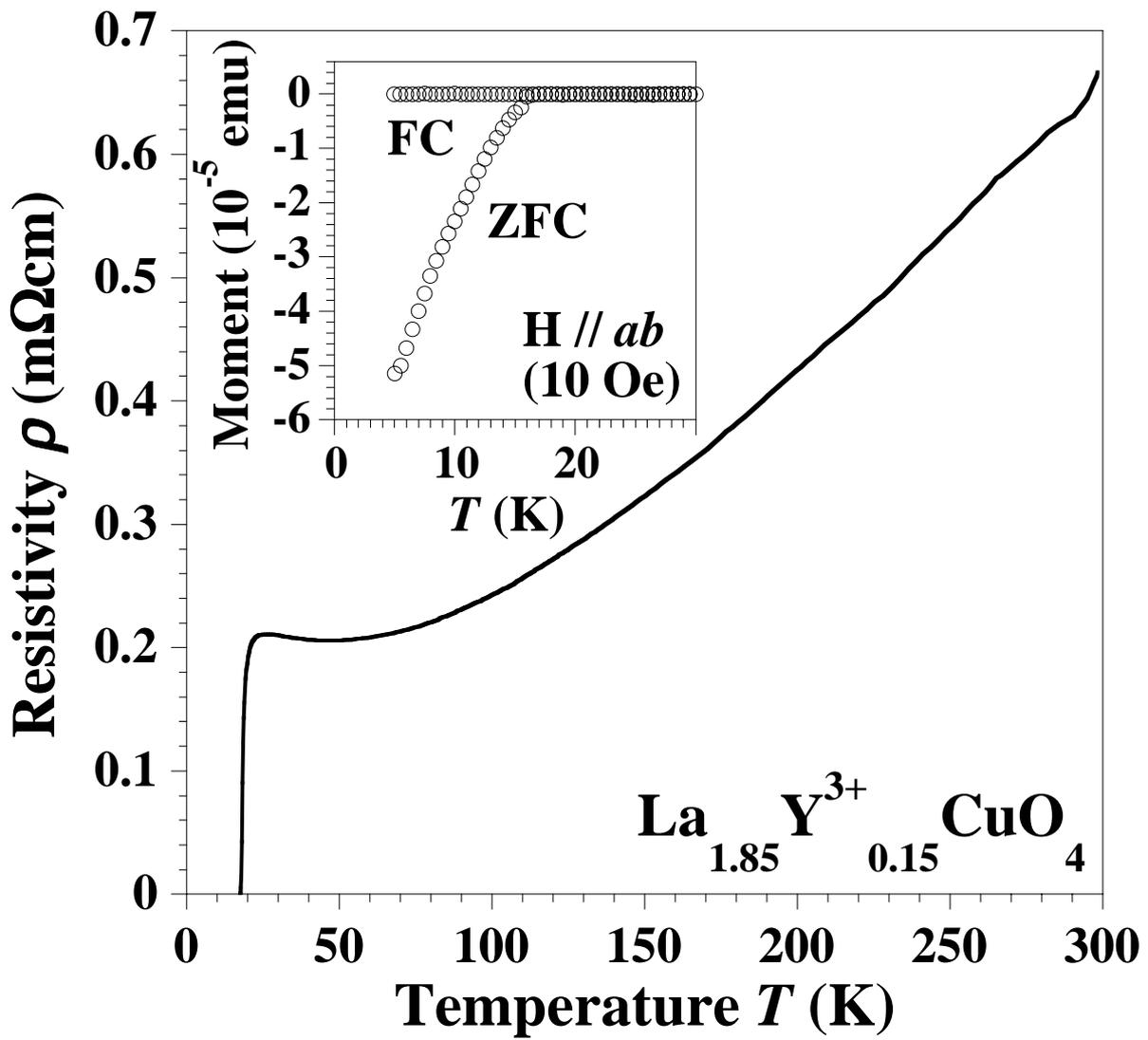

Figure. 3 A. Tsukada *et al.*

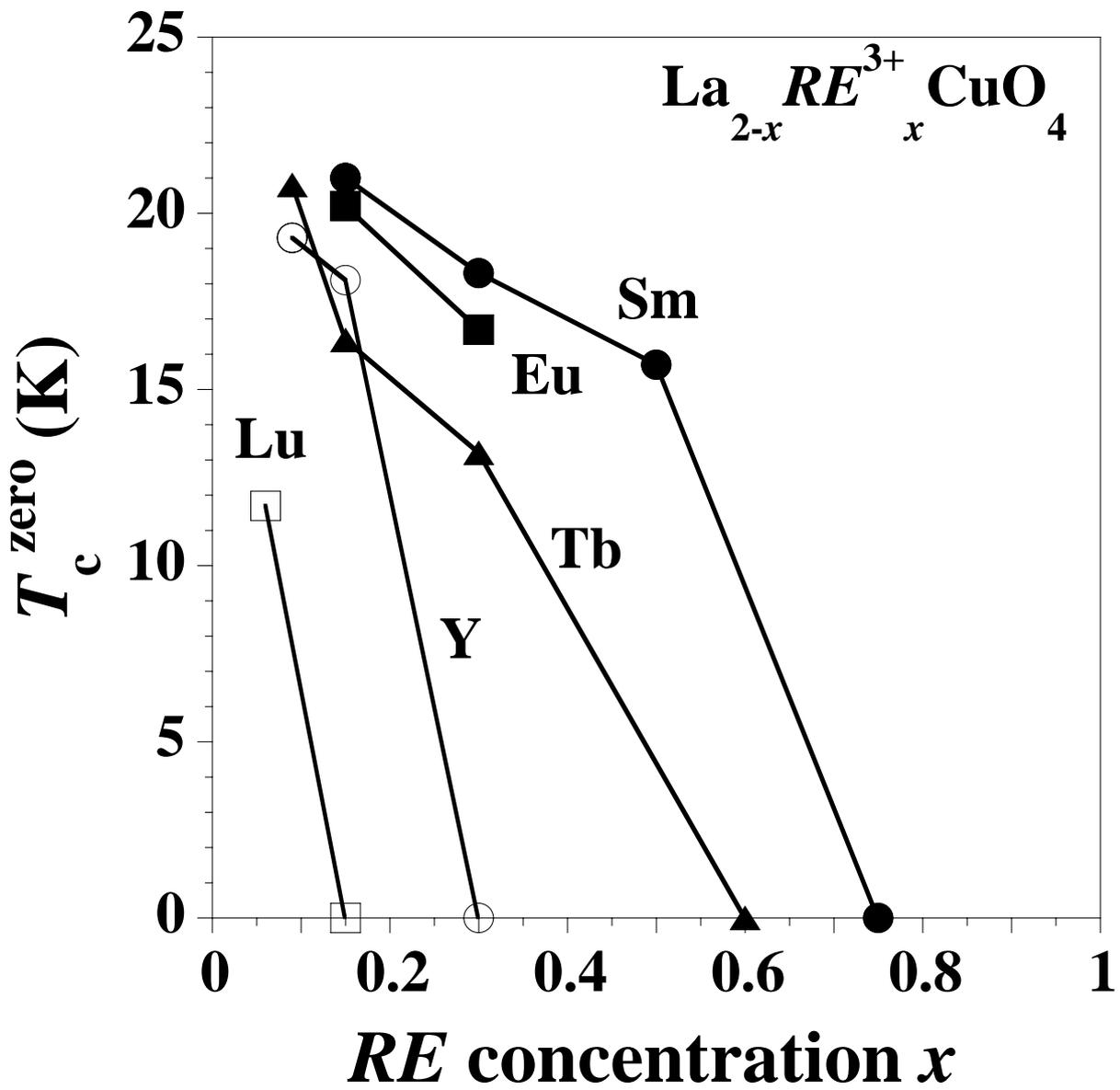

Figure. 4 A. Tsukada *et al.*